\documentclass[a4paper,11pt]{article}
\pdfoutput=1 

\usepackage{jinstpub} 

\usepackage{lineno}

\title{\boldmath First measurements with a new $\beta$-electron detector for spectral shape studies}

\author[a,b]{V.~Guadilla,}
\author[c,d]{A.~Algora,} 
\author[a]{M.~Estienne,}  
\author[a]{M.~Fallot,}  
\author[e]{W.~Gelletly,}  
\author[a]{A.~Porta,}  
\author[a]{L.-M.~Rigalleau,}
\author[a]{J.-S.~Stutzmann}

\affiliation[a]{Subatech, IMT-Atlantique, Universit\'e de Nantes, CNRS-IN2P3, F-44307, Nantes, France} 
\affiliation[b]{Faculty of Physics, University of Warsaw, 02-093 Warsaw, Poland}
\affiliation[c]{Instituto de F\'isica Corpuscular, CSIC-Universidad de Valencia, E-46071, Valencia, Spain}
\affiliation[d]{Institute of Nuclear Research of the Hungarian Academy of Sciences, Debrecen H-4026, Hungary}
\affiliation[e]{Department of Physics, University of Surrey, GU2 7XH, Guildford, UK}

\emailAdd{vguadilla@fuw.edu.pl}

\abstract{The shape of the electron spectrum emitted in $\beta$ decay carries a wealth of information about nuclear structure and fundamental physics. In spite of that, few dedicated measurements have been made of $\beta$-spectrum shapes. In this work we present a newly developed detector for $\beta$ electrons based on a telescope concept. A thick plastic scintillator is employed in coincidence with a thin silicon detector. The first measurements employing this detector have been carried out with mono-energetic electrons from the high-energy resolution electron-beam spectrometer at Bordeaux. Here we report on the good reproduction of the experimental spectra of mono-energetic electrons using Monte Carlo simulations. This is a crucial step for future experiments, where a detailed Monte Carlo characterization of the detector is needed to determine the shape of the $\beta$-electron spectra by deconvolution of the measured spectra with the response function of the detector. A chamber to contain two telescope assemblies has been designed for future $\beta$-decay experiments at the Ion Guide Isotope Separator On-Line facility in Jyv\"askyl\"a, aimed at improving our understanding of reactor antineutrino spectra.}


\arxivnumber{2305.13832} 

\begin{document}
\maketitle
\flushbottom

\section{Introduction}\label{sec-1}

The precise measurement of the different types of radiation emitted in $\beta$ decay is well established as an experimental tool to improve our knowledge of nuclear structure. The widespread spectroscopy measurements of $\beta$-delayed $\gamma$ rays is in stark contrast with the relatively few $\beta$-electron measurements that have been made. The study of $\beta$-electron spectra is of special interest because of the theoretical implications that can be extracted from their shape. The $\beta$ spectrum can be described within the Behrens-B\"uhring formalism~\cite{Behrens_beta} as:

\begin{equation}
\frac{dN_{\beta}}{dW}=pW(W-W_0)^2F(Z,W)C(Z,W)K(Z,W)
\label{beta_spec}
\end{equation}

\noindent where the total $\beta$ energy is represented by $W$, $W_0$ is the endpoint of the spectrum, Z is the atomic number of the daughter nucleus, $p$ is the momentum of the electron and $F(Z,W)$ represents the Fermi function. The $\beta$ spectrum in Equation~\ref{beta_spec} is also proportional to two terms that will be discussed in what follows: a shape factor, $C(Z, W)$, that contains nuclear structure information about the kind of transition involved, and $K(Z, W)$, that gathers together a series of correction terms including the finite size of the nuclear electric charge, screening of atomic electrons, finite size distribution of decaying neutron, quantum electrodynamics radiative corrections, weak magnetism correction etc. (for a detailed discussion of all these terms, the reader is referred to~\cite{RevModPhys_allowed,HAYEN_generator,Huber}).

Amongst all the factors that comprises $K(Z, W)$, the weak magnetism form factors are of particular interest in studies of the properties of the weak interaction and the search for non-Standard Model contributions. For example, if the scalar and tensor coupling constants lie outside the Standard Model values, then it would be reflected in the $\beta$ spectrum by means of the so-called Fierz interference term. For a review of $\beta$-shape studies related to fundamental properties of the weak interaction and the search for non-standard model components see~\cite{Severijns_2014,GONZALEZALONSO2019165}. Some recent examples of the search for exotic tensor contributions to the weak magnetism correction in Gamow-Teller (GT) $\beta$ decays 
are the investigation of the $\beta$ spectra of $^{6}$He and $^{20}$F by implanting these ions in CsI(Na) or NaI(Tl) crystals~\cite{6He_beta}, the study of the $\beta$ spectrum of $^{45}$Ca with a combination of a magnetic spectrometer and silicon detectors~\cite{Nab_45Ca} and the measurement of the $\beta$ spectrum of $^{114}$In with a plastic scintillator combined with a multi-wire drift chamber~\cite{MiniBETA_2022}. These studies require a precision in the measurement of the $\beta$-spectrum shape at the level of 0.1\%, which is beyond the scope of this work.

The $\beta$ decay shape factor, $C(Z, W)$, depends on the change in the value of the orbital angular momentum ($\Delta L$) and the parity ($\Delta\pi$) between the initial and final states. The theoretical description of the shape factors for allowed transitions ($\Delta L$=0 and $\Delta\pi$=0) has been thoroughly investigated~\cite{RevModPhys_allowed}, and a recent systematic comparison of experimental and theoretical spectra showed an excellent agreement~\cite{PRC_Mougeot2015}. The situation is more complicated for forbidden transitions ($\Delta L>$0) since the shape factor $C(Z, W)$ increases in complexity with the degree of forbiddenness of the decay. Among forbidden transitions, the shape factors for unique decays, those with orbital angular momentum and spin angular momentum aligned, are simpler than for the general case of non-unique transitions, but in both cases a lot of effort is being made to find an adequate theoretical description. A recent example is the theoretical study of two second-forbidden non-unique $\beta$ transitions by means of microscopic effective interactions and phenomenological effective interactions~\cite{PhysRevC_2020_24Na}. 

The shape factors $C(Z, W)$ can be employed to study the effective values of the weak coupling constants with the so-called spectrum-shape method (SSM)~\cite{PhysRevC_Suhonen_2016,PhysRevC_Suhonen_2017_1,PhysRevC_Suhonen_2017_2}. This method relies on the sensitivity of the theoretical $\beta$-decay shape factors to the ratio between the vector and the axial-vector coupling constants, $g_V$ and $g_A$, respectively. For the most sensitive cases, a comparison of theoretical predictions with $\beta$-spectra measurements allows one to extract information on the effective value of $g_A$. Some recent applications of the SSM method are the measurement of the fourfold forbidden non-unique decay of $^{113}$Cd using an array of CdZnTe semiconductor detectors~\cite{COBRA_2019,COBRA_2021}, and the bolometric measurement of the fourfold forbidden $\beta$ decay of $^{115}$In by means of a LiInSe$_2$ crystal~\cite{LiInse_2022}.

The description of forbidden transitions turns out to be important in reactor physics. Mean $\beta$ energies involved in the evaluation of reactor decay heat, the energy released in the radioactive decay of fission fragments, are known to be affected by the $\beta$-shape factor employed to calculate them. Errors in the mean $\beta$ values available in the nuclear databases associated with an improper treatment of forbidden transitions have already been pointed out~\cite{PRC_Mougeot2015}, and they are known to affect not only reactor decay heat calculations but also other applications such as dose rate evaluations of natural radionuclides~\cite{EPJ_40K_2019}.

The interest in a proper description of forbidden transitions in the context of reactor physics has increased in recent years due to their role in the accurate evaluation of the reactor antineutrino spectrum, given the electron-antineutrino correspondence due to energy conservation. A correct treatment of the $\beta$-decay forbidden transitions of the fission fragments could help to shed light on two questions that puzzle the neutrino community nowadays when experimental reactor antineutrino spectra are compared with predictions: the reactor antineutrino flux anomaly~\cite{Anomaly} and the spectral shape distortion observed at 4-7~MeV~\cite{DoubleChooz_bump_2015,PRL_RENO_bump,DayaBay_bump_2017,PRL_NEOS_bump}. The approximations normally introduced in the treatment of forbidden transitions for the evaluation of reactor antineutrino spectra, typically assuming allowed shapes, have been shown to be non-negligible sources of uncertainty~\cite{PhysRevLett_Hayes_2014, PhysRevC_2015_Fang, PRC_Mougeot2015,PRD_2019_conversion_china, PhysRevC_2022_92Rb}. 

Recently, shell model calculations of shape factors for the dominant forbidden transitions in the 4-7~MeV region have shown the potential to improve our understanding of both the reactor flux anomaly and the spectral shape distortion~\cite{PRC_Hayen_2019,PRC_R_Hayen_2019}. A similar conclusion has recently been reached after considering appropriately allowed and first-forbidden transitions for the calculation of the $^{92}$Rb $\beta$ spectrum~\cite{PhysRevC_2022_92Rb}, one of the main contributors to the total reactor antineutrino spectrum. These results confirm the need for a proper theoretical treatment of forbidden transitions and the importance of experimental $\beta$-spectra measurements for the most relevant nuclei contributing to the reactor antineutrino spectra, as suggested by Sonzogni \textit{et al.}~\cite{Sonzogni_2017_flux_PRL}. Such measurements have been encouraged by an International Atomic Energy Agency (IAEA) committee of experts~\cite{IAEA_2019} in order to validate and constrain theoretical predictions. In this context, the first-forbidden non-unique $\beta$ spectrum of the decay of $^{137}$Xe, important for the reactor antineutrino spectrum, was measured recently with the EXO-200 time projection chamber (TPC)~\cite{PhysRevLett_EXO_2020}, finding good agreement with shell model calculations.

In the 1980s the individual $\beta$ spectra of more than 100 fission products were measured at ISOLDE and OSIRIS by Tengblad \textit{et al.}~\cite{Olof_beta}. Two telescopes were employed for the detection of $\beta$ electrons: one for low-energy electrons (250-1500~keV) based on a thick silicon detector in coincidence with a plastic scintillator detector, and a second one for electrons between 1 and 14~MeV based on a high-purity germanium detector (HPGe) in coincidence with a plastic scintillator. Recent comparisons of these measurements with $\beta$ spectra constructed assuming allowed shapes from total absorption $\gamma$-ray spectroscopy (TAGS) results have shown some discrepancies~\cite{Val17,Simon_PRC,PRC_BDN,PRC_96Y}. Both experimental techniques are free from the Pandemonium systematic error~\cite{Har77} and the origin of such discrepancies is not understood, which calls for new $\beta$-spectra measurements of fission products to clarify them. 

In this work we will describe a new detector system aimed at providing new measurements of $\beta$-electron spectra with a precision of the order of 1-5\% in order to distinguish the shape effects introduced by forbidden transitions. The idea for the system was triggered by the above-mentioned need for a better understanding of reactor antineutrino spectra, but it could also be used for the application of the SSM method in some particular cases. A description of the detector will be presented in Section \ref{sec-2} and we will show the first measurements carried out with mono-energetic electrons in Section \ref{sec-3}. One of the main goals of these first measurements is to provide a set of reference measurements for the Monte Carlo (MC) evaluation of the response function of the detector, essential for the deconvolution of future $\beta$-spectra measurements. Thus, in Section \ref{sec-4} we will discuss the MC characterization of the detector, by comparing experimental and simulated mono-energetic spectra. Section~\ref{sec-5} will address the conclusions and future prospects. 

\section{The $\beta$-electron detector}\label{sec-2}

A variety of approaches to measuring $\beta$ spectra has been adopted over the years. Apart from those already mentioned in the introduction, some examples are the 4$\pi$ semiconductor spectrometer used by Spejewski in combination with a NaI(TI) well-shaped crystal~\cite{SPEJEWSKI1966481}, the spectrometer recently developed at LNE-LNHB using two Passivated Implanted Planar Silicon (PIPS) detectors~\cite{NIMA_2023_CEA} or the recent approach of the St. Petersburg group based on a full absorption Si(Li) detector and a transmission Si detector~\cite{NIMA_rusos_Si_2018, 210Bi_rusos_Si_2020, NIMA_2023_rusos}. Such detectors are designed to be used with $\beta$ sources sandwiched inside the detection system, in order to diminish backscattering effects. Note that in the present case, we are mainly interested in measurements of short-lived decays of fission fragments for which the production of $\beta$ sources is, in general, not feasible.   

\begin{figure}[h]
\begin{center} 
\includegraphics[width=0.7 \textwidth]{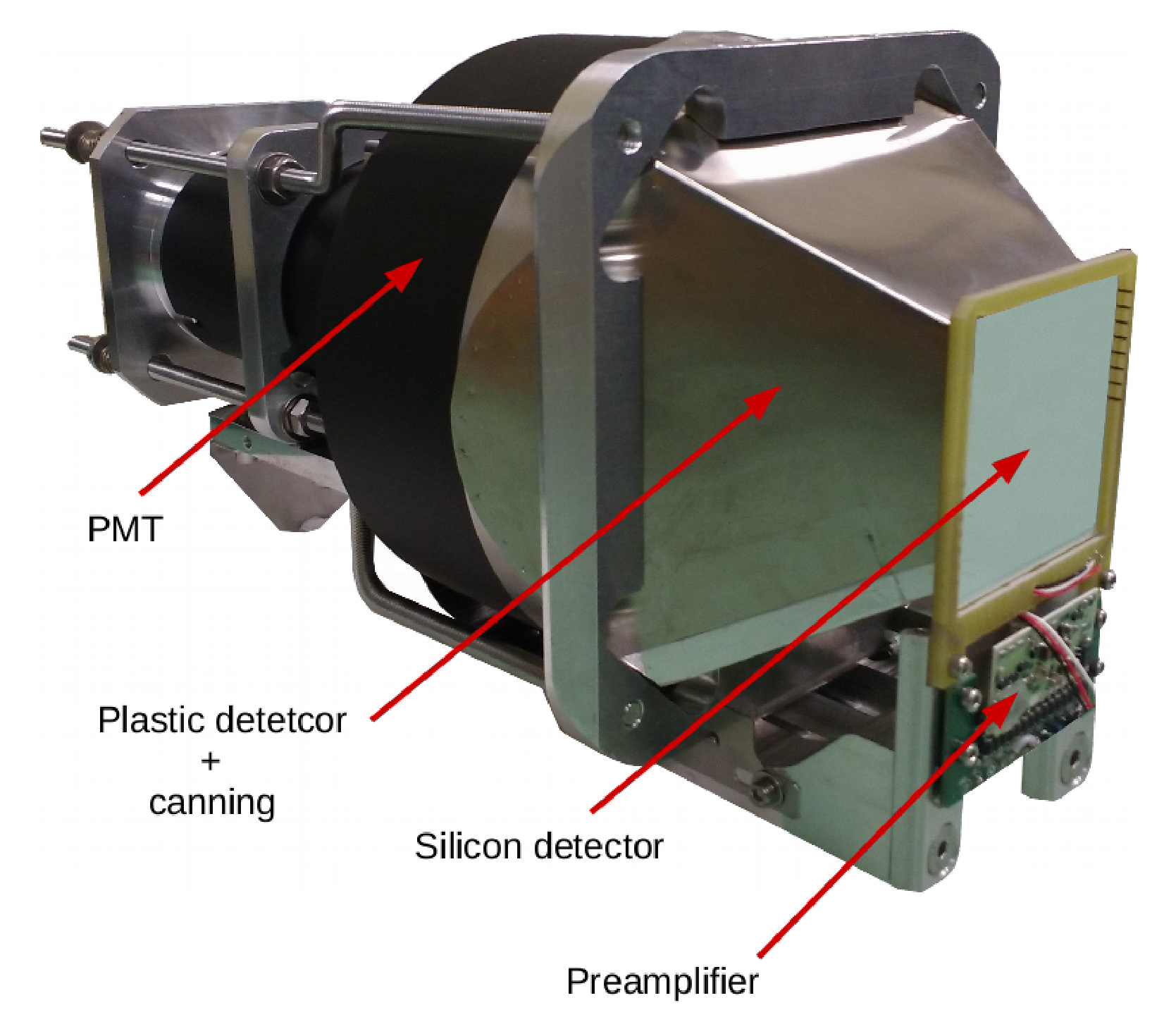}
\caption{$\Delta E$-$E$ telescope mounted on the mechanical support. The main parts of the set-up are indicated: the silicon detector connected to the preamplifier and the plastic detector surrounded by the canning and coupled to the PMT. See text for more details.}
\label{beta_det}
\end{center}
\end{figure}

The $\beta$-electron spectrometer presented here has been designed and constructed based on a silicon-plastic telescope concept~\cite{Master_Hector}, allowing an arrangement with several telescopes around an implantation point. We have chosen a similar solution to that of Horowitz \textit{et al.}~\cite{HOROWITZ1994522}, that employed a thin silicon detector in coincidence with a thick BC404 plastic scintillator. Such an approach for a $\beta$ telescope was also successully employed for $\beta$-$\nu$ angular correlations in the past~\cite{TRIUMF_beta_telescope, GANIL_beta_telescope}. In our case, as a $\Delta E$ detector we use a 500~$\mu$m-thick MSX025 non-segmented silicon detector from MICRON~\cite{micron}, with 50$\times$50~mm$^2$ active area and mounted on a PCB frame. The $E$ detector consists of a 7.5~cm-thick plastic detector made of EJ200 scintillation material~\cite{EJ200_manual}, manufactured by Scionix~\cite{scionix}. The thickness was chosen to stop $\beta$ electrons corresponding to $Q_{\beta}$ values up to 8-10~MeV, typical of some of the most relevant decays for the reactor antineutrino spectrum. The plastic material has been polished and the rounded sides painted with EJ510 reflective paint~\cite{EJ510_manual}. Thanks to a truncated cone design (see Figure~\ref{beta_det}) the shape of the plastic detector matches the 50$\times$50~mm$^2$ active area of the silicon detector at the front face, and the 133~mm-diameter circular shape of the photomultiplier tube (PMT) used for readout at the rear face. A R877-02 PMT from Hamamatsu~\cite{hamamatsu} is employed, modified to work under vacuum and surrounded by E989-26 magnetic shielding~\cite{hamamatsu}. The PMT works at -1500~V supplied by a CAEN 470 high voltage power supply~\cite{caen}. A CAEN A1422H preamplifier board with 90~mV/MeV amplification~\cite{caen} is used with the silicon detector. The preamplifier board has been modified to enable the transit of the HV line, thus working independently of a preamplifier box. The preamplifier is powered by a $\pm$12~V external module and the silicon detector is polarized through the preamplifier by a Mesytec MHV-4 bias supply~\cite{mesytec}.

The whole telescope assembly is mounted on a support made of aluminum and stainless steel, as shown in Figure~\ref{beta_det}. This support allows us to mechanically couple the plastic detector to the PMT. A thin (0.5~mm) stainless-steel canning covers the four sides of the plastic detector as shown in Figure~\ref{beta_det} in order to prevent outside light from leaking inside and protect the plastic from physical damage when handling the telescope.

\begin{figure}[h]
\begin{center} 
\includegraphics[width=0.5 \textwidth]{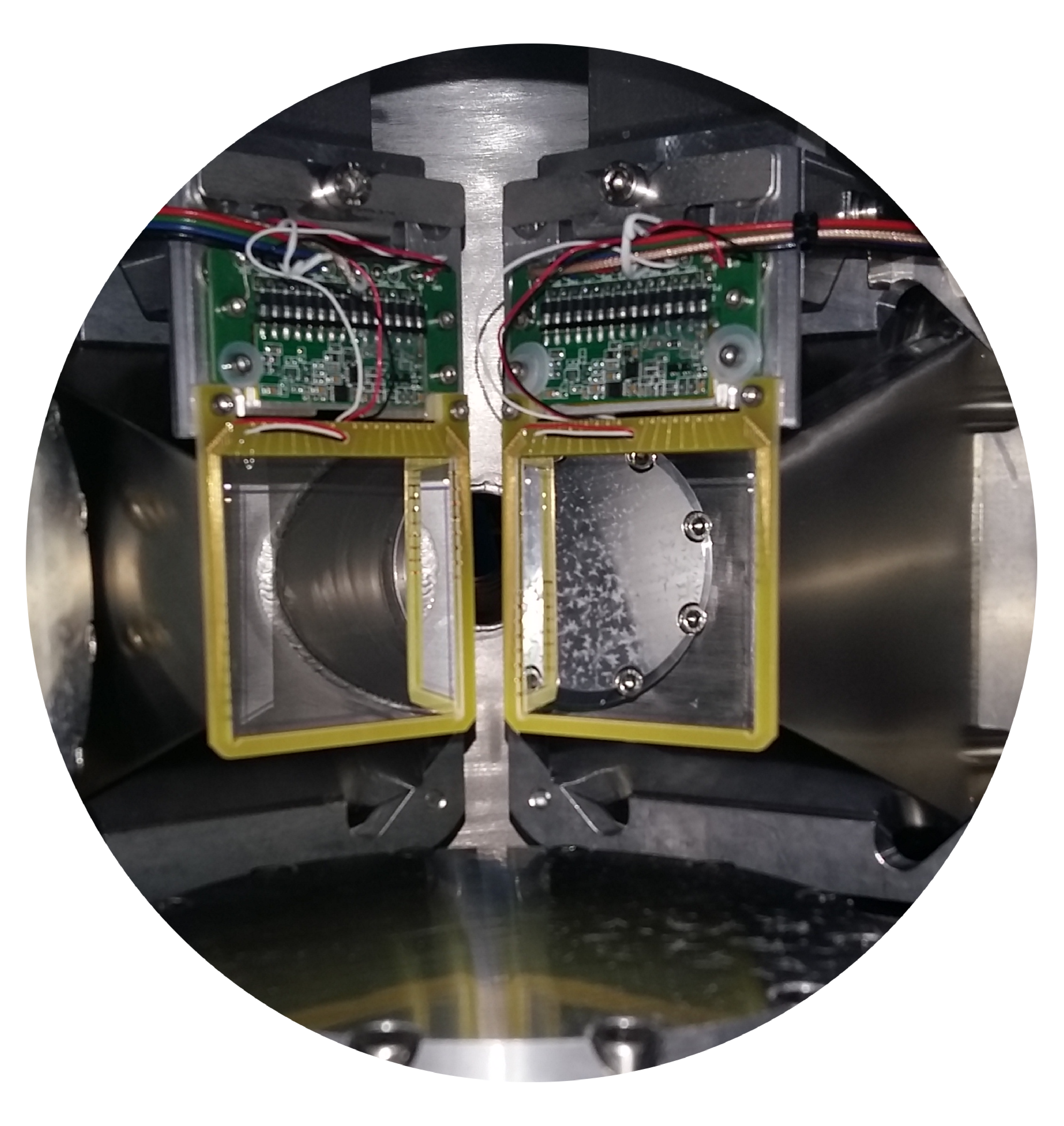}
\caption{Two $\Delta E$-$E$ telescopes inside a vacuum chamber designed for future experiments. They are placed symmetrically with respect to a movable magnetic tape (not shown in the picture for clarity). See text for further details.}
\label{beta_chamber}
\end{center}
\end{figure}

A vacuum chamber to contain two telescope assemblies has been designed at Subatech for future experiments at the Ion Guide Isotope Separator On-Line (IGISOL) facility of the University of Jyv\"askyl\"a~\cite{Moore_IGISOLIV}. The front faces of the two detectors are placed at an angle to one another. The lines from the centers of the two faces form an angle of 70$^{\circ}$ (see Figure~\ref{beta_chamber}) and they are placed symmetrically with respect to the point where a movable magnetic tape will be located, at 5~cm from the front faces of the silicon detectors. This tape will be employed for implantation of the nuclei of interest and the subsequent removal of the remaining activity in cycles that will depend on the half-life of the nuclei involved. The design of the detector has been optimized for this configuration, minimizing the amount of dead material while using vacuum compatible elements. Thanks to its truncated-conical shape the plastic detector covers 132$^{\circ}$ out of the 140$^{\circ}$ subtended by the two telescopes from the center of the magnetic tape.

\begin{figure}[h]
\begin{center} 
\includegraphics[width=0.8 \textwidth]{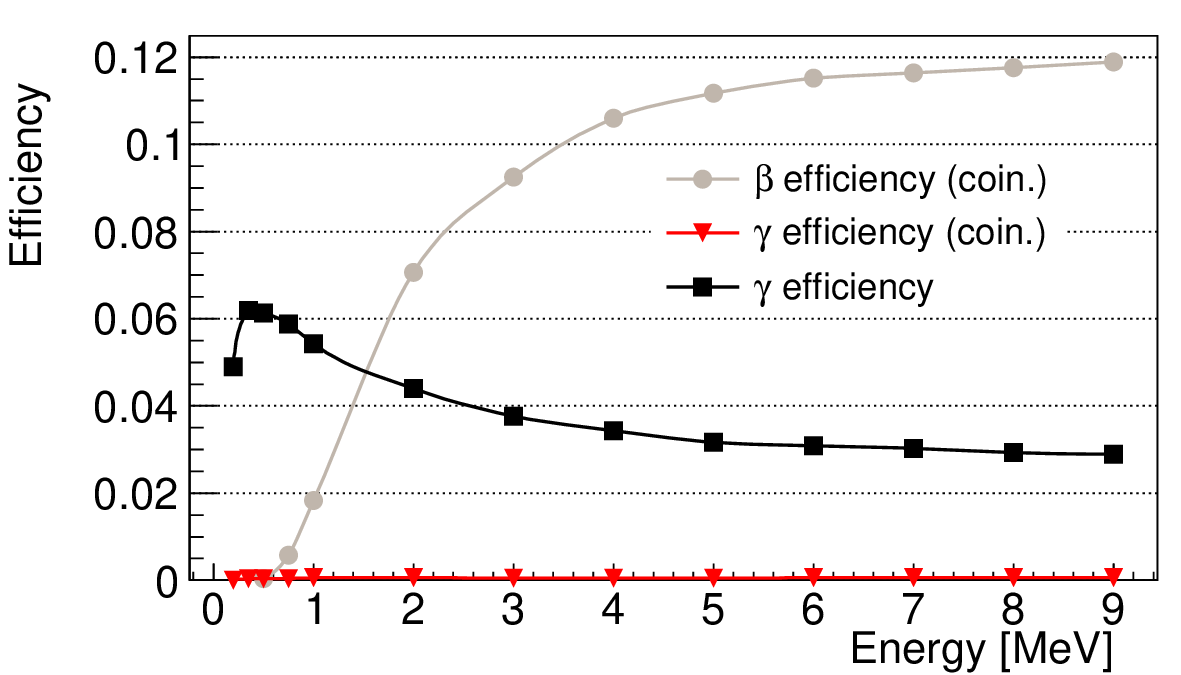}
\caption{Simulated efficiency of the setup with two telescopes for $\beta$ electrons (gray line with circular dots) and $\gamma$ rays (red line with triangular dots). The simulated efficiency of the assembly corresponding to the plastic detectors not in coincidence with the corresponding silicon $\Delta E$ detectors is also included (black line with square dots) to illustrate the $\gamma$-suppression effect of the $\Delta E$-$E$ coincidence.}
\label{eff}
\end{center}
\end{figure}

We have calculated the total efficiency of the two telescopes for $\beta$ particles by means of MC simulations, that are detailed in Section~\ref{sec-3} and presented in Figure~\ref{eff}. It ranges from 2$\%$ for a 1~MeV $\beta$ end-point to more than 10$\%$ for $\beta$ end-points above 3~MeV. The effectiveness of the silicon-plastic coincidence to reduce the sensitivity to $\gamma$ rays is proved by the $<$0.1$\%$ total efficiency of the silicon-plastic coincidence for 500~keV $\gamma$ rays, in contrast with the 6$\%$ value for the plastic detector when it is not in coincidence with the silicon detector. 

\section{Characterization of the response function of the detector} 

In order to determine the $\beta$-decay electron spectrum (\textbf{\textit{b}}) from measurements with our $\Delta E$-$E$ telescopes, one possibility is to perform the deconvolution of the measured $\beta$-electron spectrum (\textbf{\textit{d}}) using the response function of the detector (\textbf{\textit{R}}). This is a similar inverse problem to that solved in TAGS analyses~\cite{TAS_algorithms}: \textbf{\textit{d=R} $\otimes$ \textit{b}}. However, in the TAGS technique the response function of the total absorption $\gamma$-ray spectrometer combines $\gamma$ rays and $\beta$ particles (and eventually X rays and conversion electrons), being thus dependent on the branching ratios for the different de-excitation paths of the states populated in the decay~\cite{TAS_MC}. In our case, the response function of the $\Delta E$-$E$ detector will not depend on the details of the decay, and it will just consist of the ensemble of responses of the detector to mono-energetic electrons of different energies. As in the TAGS technique~\cite{TAS_MC}, this response matrix will be calculated by means of MC simulations, properly validated with reference measurements, as we present in the following subsections.

\subsection{Measurements with mono-energetic electrons}\label{sec-3}

The characterization of the response function of a $\beta$ detector to mono-energetic electrons is constrained by the limited availability of mono-energetic electron sources. For calibration, typically the conversion electrons of a $^{207}$Bi source are employed, with the main electron lines coming from the internal conversion of transitions of 569.7, 1063.6 and 1770.2~keV in $^{207}$Pb~\cite{FUJITA_207Bi}. $E0$ transitions are other sources of mono-energetic electrons. One possibility is to measure them as part of the de-excitation of the daughter nucleus levels populated in $\beta$ decay. However, it is difficult to find feasible cases that could be easily produced in a radioactive beam facility and for which levels de-exciting through $E0$ transitions are strongly populated. In addition, the majority of such transitions are limited to energies below 2~MeV. The reader will find a review of known $E0$ transitions in Ref.~\cite{E0_review}. 
Alternatively, one can think of using electron beams from electron linear accelerators. However, it is very difficult to get the proper energy range, 0-10~MeV in this case, appropriate energy resolution and reasonable intensities at the same time. It should be noted that in our case only intensities of the order of fA are suitable to work without excessively distorting the measured spectra as a consequence of the pileup effects due to high counting rates. 

\begin{figure}[h]
\begin{center} 
\begin{tabular}{ccc}


\includegraphics[width=0.45 \textwidth]{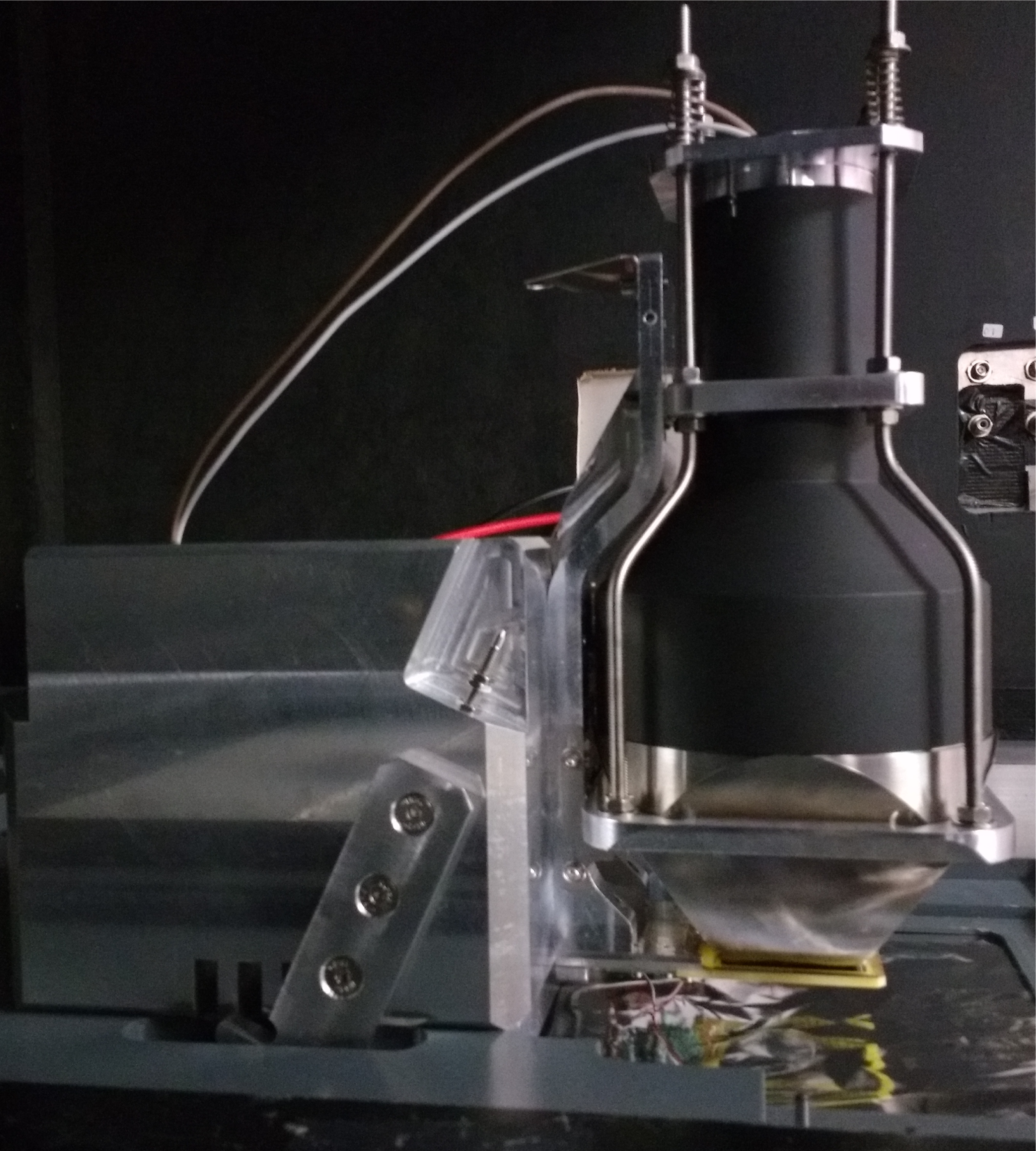} &

\includegraphics[width=0.025 \textwidth]{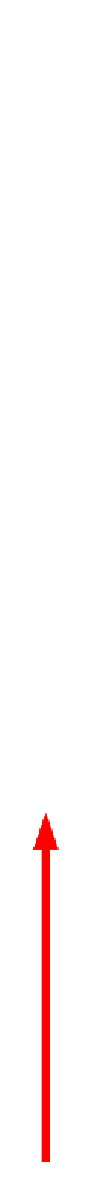} &

\includegraphics[width=0.45 \textwidth]{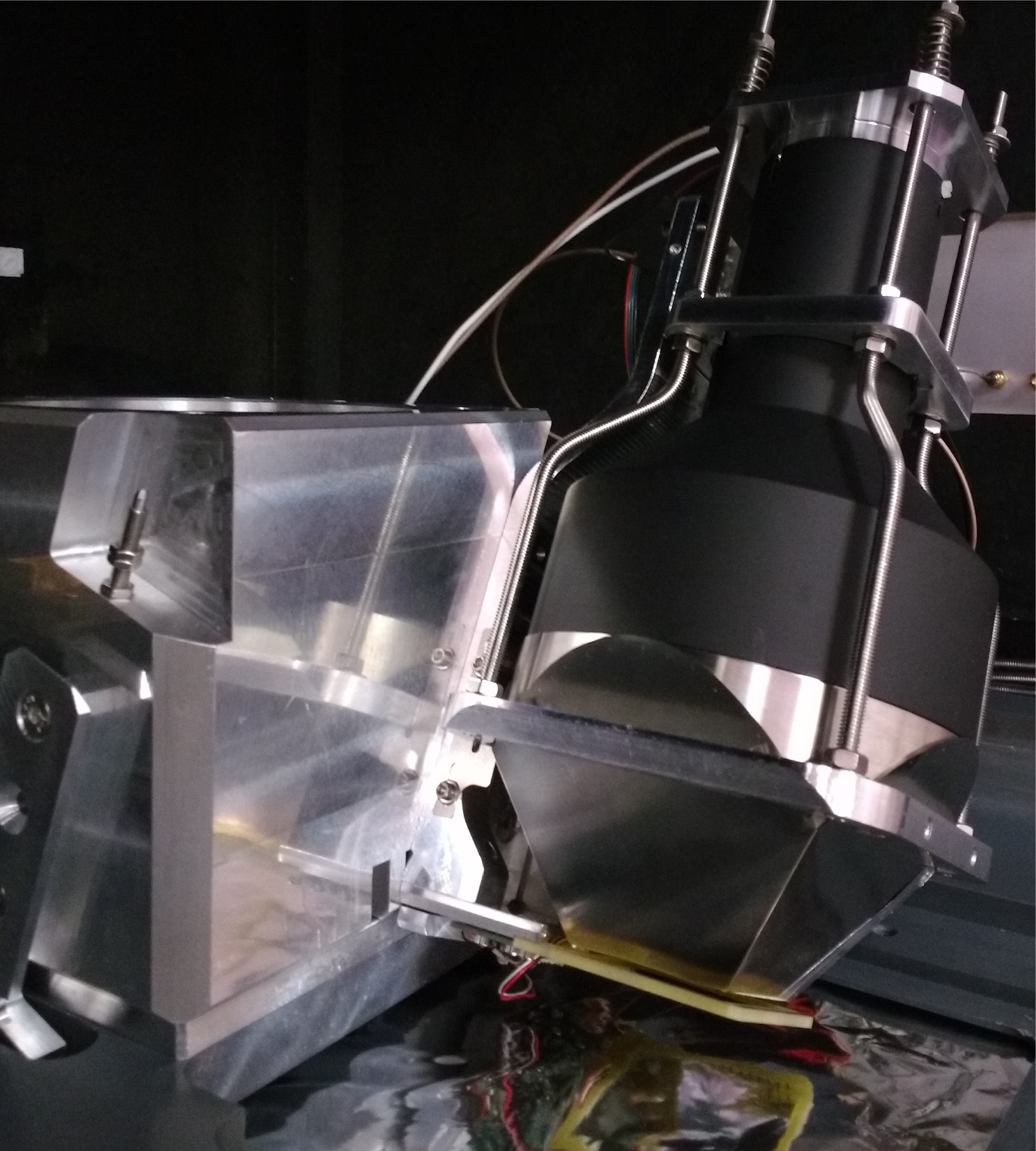} 

\end{tabular}
\caption{$\Delta E$-$E$ telescope inside the light-tight box coupled to the high-energy resolution electron beam spectrometer at CENBG~\cite{Marquet}. The electron-beam direction is represented by a red arrow. The telescope at 0 degrees with respect to the beam direction is shown in the left hand, while the telescope rotated 20 degrees can be seen in the picture on the right.}
\label{setup_Bordeaux}
\end{center}
\end{figure}

The high-energy resolution electron-beam spectrometer at Centre d'Etudes Nucl\'eaires de Bordeaux Gradignan (CENBG)~\cite{Marquet} is a good possibility for the characterization of a $\beta$ detector with mono-energetic electrons and we have made use of it in the first measurements with the new $\beta$ detector presented in this work. The electron-beam spectrometer consists of a collimated $^{90}$Sr-$^{90}$Y source with an adjustable magnetic field properly calibrated to select the required energy of the electrons. The beam of mono-energetic electrons is delivered through a mylar window into a light-tight box in air where we have placed one telescope assembly perpendicularly to the beam direction, as shown in Figure~\ref{setup_Bordeaux}. A support for the telescope assembly was designed in order to place the front face of the silicon detector in the position where the energy calibration of the electron beam was performed. In addition, the support also allowed us to rotate the whole telescope 10, 20 and 25 degrees with respect to the beam direction. The position of the detector for each angle was adjusted to mimic the interacting conditions for the electron-detector that we will have inside the vacuum chamber shown in Figure~\ref{beta_chamber}. In this way, for a beam perpendicular to the detector (as in Figure~\ref{setup_Bordeaux} top) electrons impinge on the center of the telescope, while for a beam at 80, 70 and 65 degrees with respect to the front face of the silicon detector, i.e. the whole telescope rotated 10, 20 and 25 degrees with respect to the beam direction as in the example shown in the right hand of Figure~\ref{setup_Bordeaux}, electrons impinge on positions that are progressively closer to the sides of the face of the silicon detector, as if they were coming from the magnetic tape in the vacuum chamber. 

The triggerless FASTER data acquisition system~\cite{faster} was used for the measurements, employing 12-bit CARAS daughter boards. A charge to digital converter (QDC) algorithm was applied to the plastic detector signal coming from the PMT, while for the preamplifier signal of the silicon detector we employed a trapezoidal filter. 

\begin{figure}[h!]
\begin{center} 
\includegraphics[width=0.8 \textwidth]{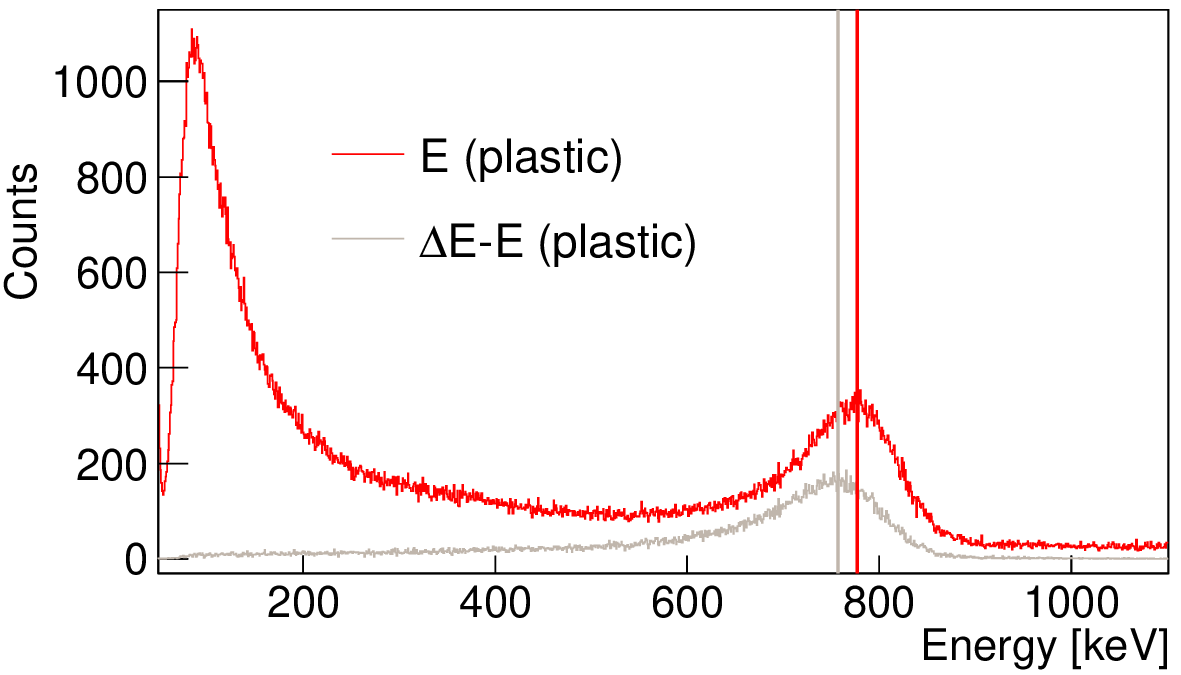}
\caption{Experimental spectrum in the plastic detector for a 1~MeV electron beam without any coincidence (in red) and in coincidence with the $\Delta E$ silicon detector (in gray). Apart from rejecting the environmental background (clearly seen at low energies), the coincidence shifts the centroid of the peak associated with the mono-energetic electrons, by about 20~keV in this case, and affects the number of counts, due to the high energy threshold of the silicon detector (see text for details). Vertical lines represent the centroids of the peaks.}
\label{singles_coin}
\end{center}
\end{figure}

\begin{figure}[h!]
\begin{center} 
\includegraphics[width=0.8 \textwidth]{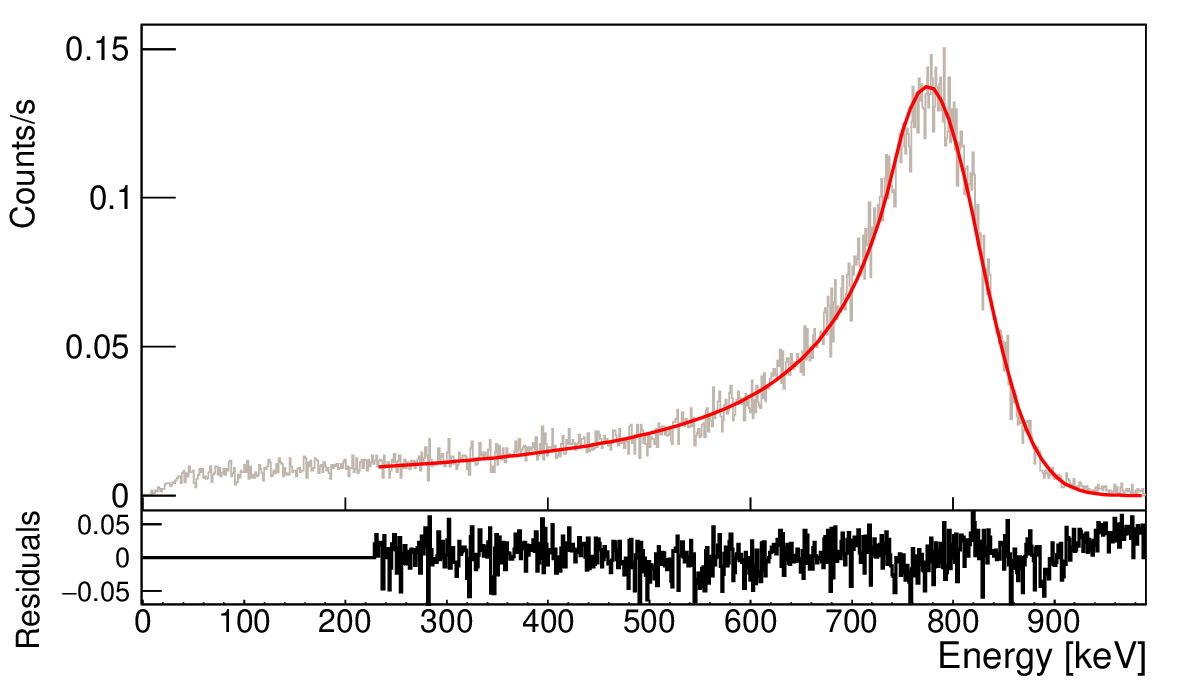}
\caption{Example of the fit (in red) of the experimental plastic spectrum (in gray) in coincidence with the silicon detector for a 1~MeV electron beam. The residuals are shown below in the fitting region, defined as the deviation between experimental and fitted spectra over the root square of the experimental number of counts for each experimental bin.}
\label{fit}
\end{center}
\end{figure}

The useful range of the spectrometer is 0.4-1.8~MeV, considering the $\beta$-energy window of the decays of $^{90}$Sr and $^{90}$Y, 0.546 and 2.282~MeV, respectively, the shape associated with these decays and the counting rate of the source~\cite{Marquet}. We measured mono-energetic electrons in this range from 0.6~MeV, because of the limited efficiency of our setup below this energy. The coincidence $\Delta E$-$E$ spectra were constructed with a timing resolution of 30~ns between plastic and silicon detectors. Electrons were fully stopped in the plastic detector, thus producing the expected mono-energetic peaks in the spectra, as shown in Figures~\ref{singles_coin} and \ref{fit} for a 1~MeV electron beam. In addition, the availability of a $^{207}$Bi source at CENBG, with well characterized geometry, allowed us to measure the spectrum of mono-energetic conversion electrons emitted. 

\begin{figure}[h!]
\begin{center} 
\includegraphics[width=0.8 \textwidth]{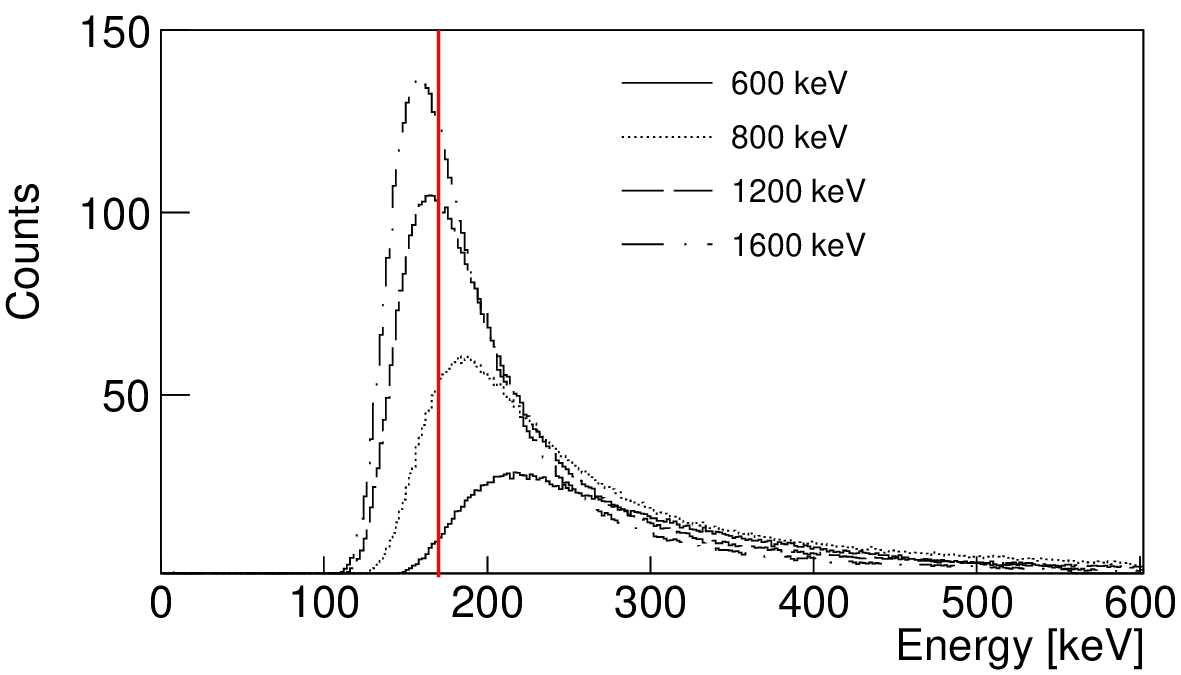}
\caption{Simulated silicon spectra for different end-point energies showing the shift to lower energies of the energy-loss peak with increasing energy (see text for details). The vertical red line represents our effective experimental threshold.}
\label{si_thres}
\end{center}
\end{figure}

The number of coincidences observed in the plastic detector is affected by the energy threshold in the silicon detector. In particular, a threshold value in the region of the energy-loss peak in the silicon detector partially cuts the full-energy electron peak in the coincidence plastic detector spectrum, thus shifting its centroid, as shown in Figure~\ref{singles_coin}. In normal conditions we typically have an energy threshold below 100~keV, but for these measurements a large effective threshold of 170~keV was observed in the silicon detector, due to grounding problems between our system and the light-tight box, and to the induced noise coming from the controllers of the spectrometer. Note that for a 500~$\mu$m-thick silicon detector, as shown in Figure~\ref{si_thres}, the energy-loss peak is centered at 150~keV for incoming electron energies above 1.5~MeV, while below this value the energy-loss peak is found at higher energies. With 400~keV electrons its value is centered around 340~keV. It implies that the effect of the threshold of the silicon detector is less pronounced for low-energy measurements. Note that for such low-energy electrons most of the energy is deposited in the silicon detector and much less energy is deposited in the plastic detector.

Due to the shift in the position of the coincidence peaks, the energy calibration of the plastic detector was performed without coincidences with the silicon detector. The value of the energy deposited by the mono-energetic beam in the plastic detector was taken from MC simulations. The details about the simulations are presented in the next section. The effective threshold of the silicon detector was then inferred by matching measurements and simulations of the plastic spectra in coincidence with the silicon detector.

The full-energy electron peaks, both in experimental and simulated spectra, were fitted with a Gaussian function with a low energy power-law tail, the so-called Crystal Ball function (CB) developed within the Crystal Ball Collaboration~\cite{PhysRevLett_CristalBall}. An example of the fit of a 1~MeV electron beam spectrum is presented in Figure~\ref{fit}. An energy resolution of 15$\%$ is obtained for this case with the $\Delta E$-$E$ coincidence spectrum.

\subsection{Monte Carlo simulations}\label{sec-4}

The simulation of electrons is considered a complex problem, mainly due to the difficulties in modeling and validating electron scattering. Modern simulation codes, such as the Geant4 toolkit~\cite{GEANT4} employed in this work (version 10.05), are continuously tested and bench-marked in order to improve the simulation of electron interactions. Recent examples of validations of Geant4 simulations with experimental data for bremsstrahlung produced by electrons~\cite{NIMA_brem_2015} and electron backscattering~\cite{IEEE_back_2016,DONDERO201818,SOTI201311,G4med} show a reasonable agreement down to 100~keV, where discrepancies are found to be significant, partly due to the uncertainty of experimental data. Such a validated performance allows the community to rely on simulations for the calculation of the response function of the present $\beta$ detectors. This was not possible in the 1980s, when limitations in the simulations caused Tengblad \textit{et al.} to use a parametric response function experimentally determined with mono-energetic electrons for the deconvolution of experimental data~\cite{Olof_beta_response}. 

\begin{figure}[h!]
\begin{center} 
\includegraphics[width=0.8 \textwidth]{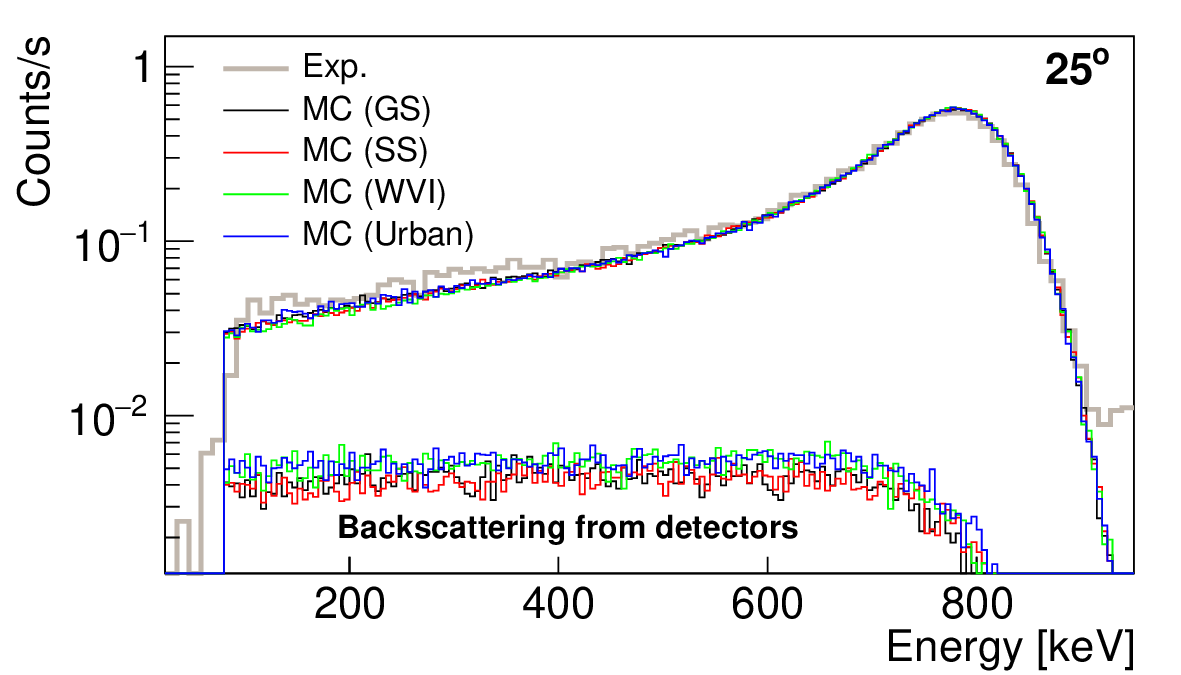}
\caption{Experimental 1~MeV electron spectra with 25$^o$ angle between the $\beta$ telescope and the direction of the electron beam (in gray). MC simulations with different scattering models are shown in different colors, normalized to the experimental spectra. The associated simulated backscattering spectra from the sensitive areas of the $\beta$ telescope are also presented.}
\label{back}
\end{center}
\end{figure}

The possible influence of different electron scattering models in our simulations has been checked by using different electromagnetic constructors, as in the most recent benchmark study of Geant4 for low-energy physics~\cite{G4med}, performed for the same version as the one used in the present work (10.05). We have considered the following electromagnetic constructors: \textit{G4EmLivermorePhysics} using the Goudsmit-Saunderson multiple scattering model (GS), \textit{G4EmStandardPhysicsSS} using the Single Scattering model (SS), \textit{G4EmStandardPhysicsWVI} using the Wentzel-VI combined multiple and single scattering model (WVI) and \textit{G4EmStandardPhysics} using the Urban electron multiple-scattering model (Urban). The reader is referred to Ref.~\cite{Ivanchenko_2010} for more information about these models. In Refs.~\cite{IEEE_back_2016,DONDERO201818, SOTI201311, G4med} the computationally demanding SS model was found to be more accurate when compared with backscattering experimental data. In our case, however, no significant influence was found on our energy-deposited spectra, as shown in Figure~\ref{back}, where we compare the experimental spectrum of 1~MeV electrons impinging on the telescope assembly at 25$^o$ with simulations using all these electromagnetic constructors. For each case, we have also isolated those events where electrons are scattered from the sensitive volumes, i.e. the active part of the silicon detector and/or plastic scintillator, into other materials, as can be seen in Figure~\ref{back}. They amount to about 2\% of the total spectrum for all constructors, being, however, almost 30\% larger for the WVI and Urban models than for the GS and SS ones. In the light of these results, it seems clear that with our energy-deposited experimental spectra any benchmark of models is beyond our sensitivity and all of them provide us with simulations reproducing the experiment very similarly. Since our main goal is to reproduce the energy-deposited spectra for the calculation of the response function of the detector, we have finally chosen the low-energy \textit{G4EmLivermorePhysics} electromagnetic constructor for all simulations in this work.

\begin{figure}[h!]
\begin{center} 
\begin{tabular}{cc}
\includegraphics[width=0.5 \textwidth]{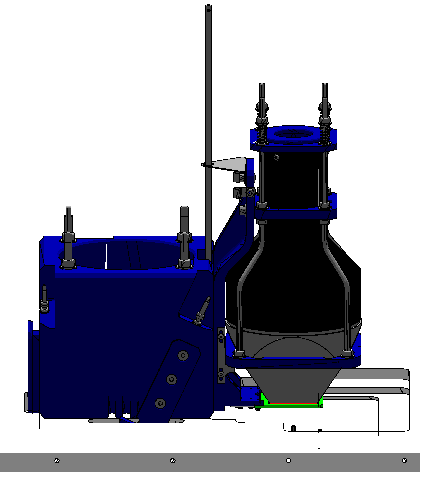} &

\includegraphics[width=0.5 \textwidth]{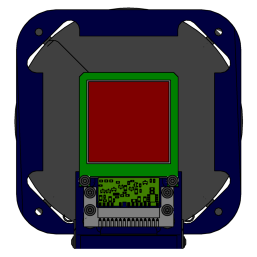}
\end{tabular}
\caption{Implementation of the geometry of the $\Delta E$-$E$ telescope in the Geant4 MC simulations. A global view of the set-up in the configuration used at CENBG is shown in the left hand (see Figure~\ref{setup_Bordeaux} for comparison), while the detailed geometries of the silicon detector and the preamplifier can be seen in the view of the front of the telescope presented in the right hand.}
\label{setup_MC}
\end{center}
\end{figure}

In our MC simulations, the geometries of the mechanical supports and the silicon preamplifier have been implemented from CATIA~\cite{catia} files by means of the CADMesh libraries~\cite{poole2012acad,poole2012fast}, that allowed us to directly read the files of the mechanical design in standard triangle language (STL) format. We have used a different input file for each material in order to customize properly the physical attributes for each volume in the Geant4 geometry description. The active parts of the set-up, i.e. the plastic and silicon detectors, have been manually implemented in the code. In Figure \ref{setup_MC} a view of the implementation of the geometry in the MC code is shown, demonstrating the level of detail achieved thanks to the CADMesh libraries, important to account for all possible materials interacting with electrons.

\begin{figure}[h!]
\begin{center} 
\includegraphics[width=1 \textwidth]{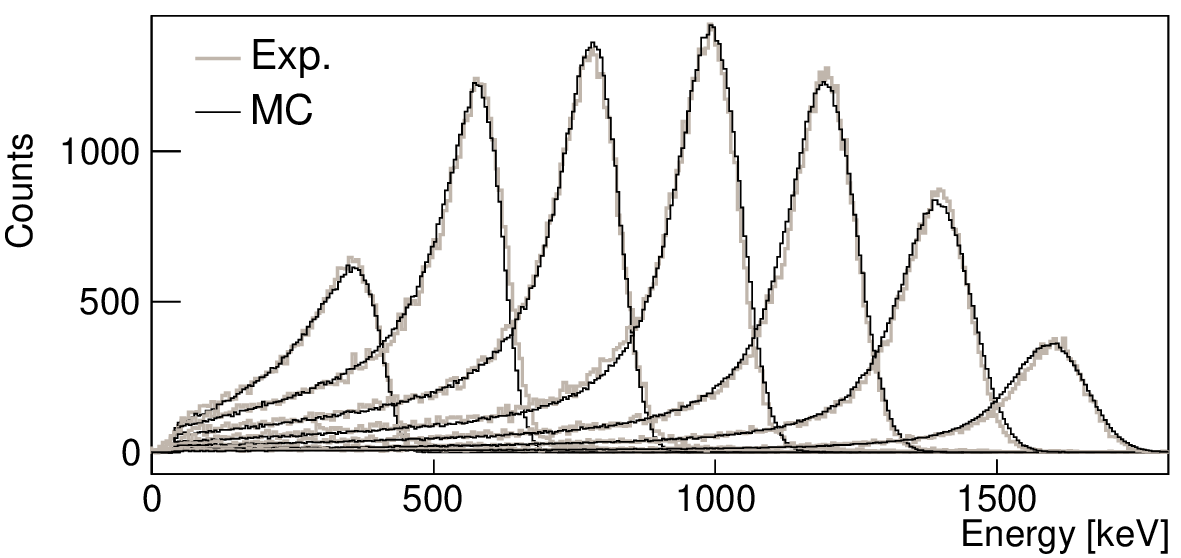}
\caption{The experimental spectra from the plastic detector in coincidence with the silicon detector (in gray) are compared with MC simulations of energy deposited (in black) for different electron beam energies from 600 to 1800~keV, in intervals of 200~keV.}
\label{energy_spectra}
\end{center}
\end{figure}

\begin{figure}[h!]
\begin{center} 
\includegraphics[width=0.8 \textwidth]{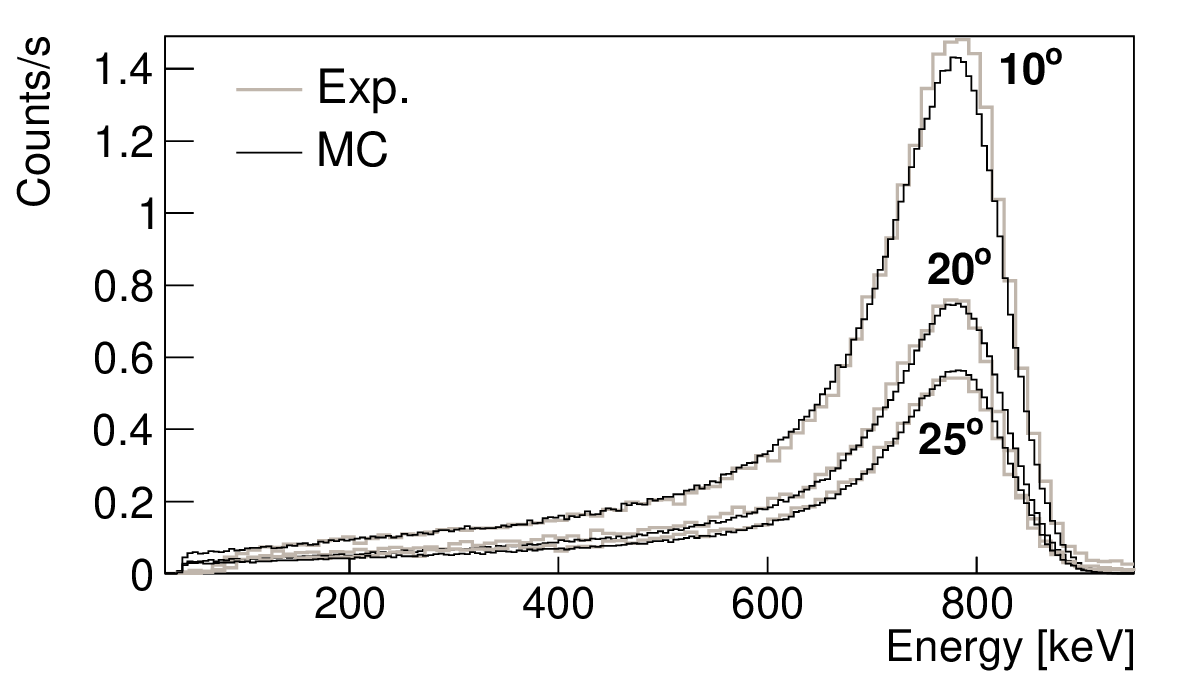}
\caption{Comparison of 1~MeV electron spectra (in gray) for different angles of the $\beta$ telescope with respect to the direction of the electron beam (in counts/s). The corresponding MC simulations are shown in black normalized to the experimental spectra. For a proper comparison of the counting rates it should be noted that here spectra are rebinned by a factor of ten with respect to the 0$^o$ spectrum shown in Figure~\ref{fit}.}
\label{angles_spectra}
\end{center}
\end{figure}

The measurements of mono-energetic electrons in the range 0.6-1.8~MeV have been compared with MC simulations. The electron beam has been implemented following the characteristics reported in~\cite{Marquet}, i.e. Gaussian profile with energy resolution of 10~keV and spatial resolution of 3~mm. For the comparison, the MC simulated spectra have been widened with an experimentally determined resolution as a function of the energy. An overall reasonable agreement is found in the shape of the peaks and in the relative tail-to-peak ratio, as can be seen in Figure~\ref{energy_spectra}. The ratio of the integral number of counts in the two spectra is used to match the MC simulations to the experimental data.  Measurements for 1~MeV electrons impinging on the telescope assembly at different angles have also been compared with simulations. The simulated geometry has been adapted accordingly to reproduce the experimental arrangement described in the previous section. In Figure~\ref{angles_spectra} we show the good agreement between experimental and simulated spectra in the shape of the peaks and in the relative tail-to-peak ratio.

\begin{figure}[h!]
\begin{center} 
\includegraphics[width=0.8 \textwidth]{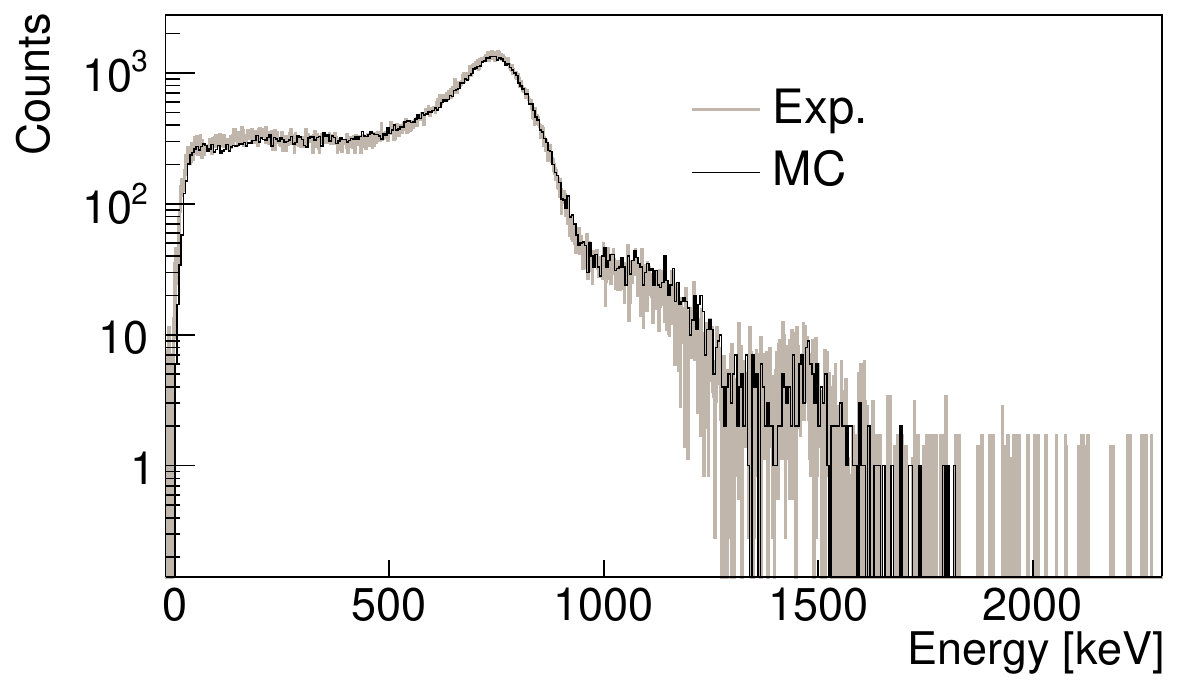}
\caption{Experimental spectrum in the plastic detector in coincidence with the silicon detector for the $^{207}$Bi source (in gray). A MC simulation of the energy deposited from this source (in black) is shown for comparison.}
\label{207Bi}
\end{center}
\end{figure}

Finally, we have also studied the reproduction of the experimental spectrum of the $^{207}$Bi source with MC simulations. The radioactive source employed is well characterized by the local group at CENBG and we have used their Geant4 implementation. It consists of an external plastic frame, an internal copper frame and a 20~$\mu$m mylar foil covering the active area. The comparison of the experimental spectrum and the simulated one is shown in Figure~\ref{207Bi}. The MC simulation reproduces the main peak observed in the measured spectrum, corresponding to the conversion electrons of the 1063.6~keV transition in $^{207}$Pb, and the weak high-energy peak corresponding to the internal conversion of the 1770.2~keV transition. Conversion electrons associated with the 569.7~keV transition in $^{207}$Pb are hardly seen at around 200~keV in the coincidence spectrum due to the large fraction of energy lost in the silicon detector.
 
\subsection{Response function simulation}\label{sec-4}

In order to ilustrate the potential performance of the present $\beta$ spectrometer, we have calculated the response function of the full assembly, with two $\Delta E$-$E$ telescopes inside the vacuum chamber presented in Section~\ref{sec-2}, for one interesting case for reactor applications: the decay of $^{94}$Y, with a $\beta$ energy window $Q_{\beta}$ of 4918(6)~keV~\cite{Qval_NNDC} and dominated by two first-forbidden transitions: a $2^-\rightarrow0^+$ ground state to ground state transition and a $2^-\rightarrow2^+$ transition populating the first excited state in $^{94}$Zr at 918.75~keV. 

For the simulation of the response function of the $\beta$ spectrometer to the $\beta$ decay of $^{94}$Y, we have once again used the CADMesh libraries~\cite{poole2012acad,poole2012fast} to implement the geometry of the chamber and of the mechanical supports from CATIA~\cite{catia} files. We have applied the same threshold conditions reported in Section~\ref{sec-3} and the same experimental width calibration determined from the measurements at CENBG to widen the simulated responses. 

\begin{figure}[h!]
\begin{center} 
\includegraphics[width=0.8 \textwidth]{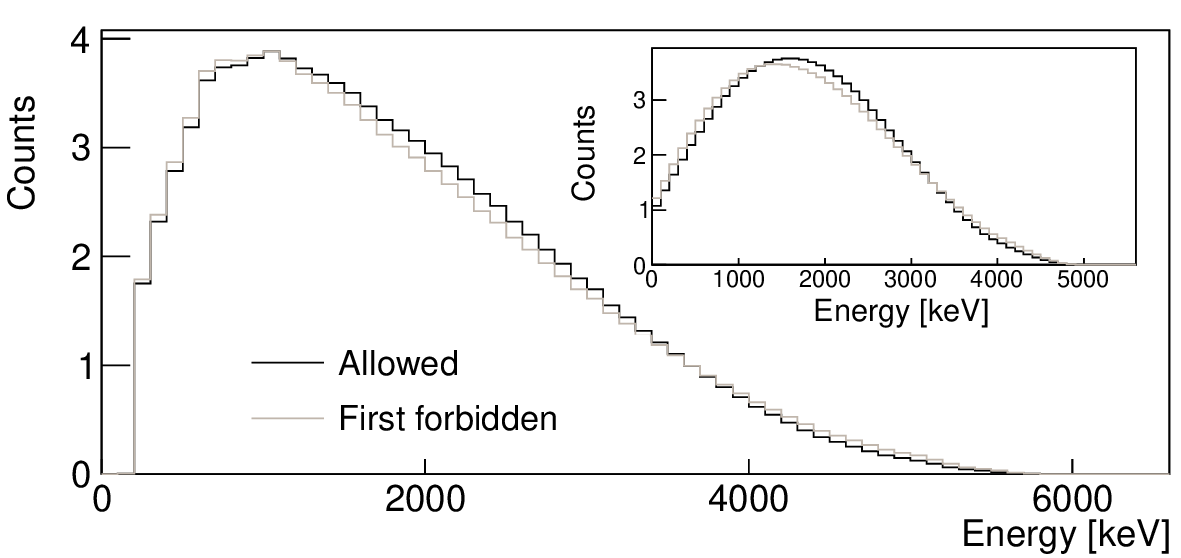}
\caption{Simulated energy-deposited $\beta$ spectrum with two $\Delta E$-$E$ telescopes for the decay of $^{94}$Y. Two theoretical $\beta$ spectra (presented in the inset) have been convoluted with the response function of the spectrometer. Spectra in black correspond to allowed shapes for all $\beta$ branches of the decay, whereas spectra in gray are associated with a first-forbidden shape treatment of the two dominant transitions. All spectra are presented with a 100~keV binning.}
\label{94Y}
\end{center}
\end{figure}

The reference theoretical $\beta$ spectrum has been calculated by means of the subroutines from the the log\textit{ft} program of NNDC~\cite{logftNNDC} assuming allowed shapes for all $\beta$ branches, with $C(Z,W)\approx1$ in Equation~\ref{beta_spec}. We have used the evaluated decay data from the ENSDF database~\cite{NDS_94}, with 41(4)\% and 39.6(22)\% $\beta$ intensities to the $0^+$ ground state and to the $2^+$ first excited state in $^{94}$Zr, respectively. 
A second theoretical spectrum has been obtained by multiplying the $\beta$ spectra of these two dominant first-forbidden branches by the shape factors $C(Z,W)$ calculated by Hayen et al.~\cite{PRC_R_Hayen_2019, PRC_Hayen_2019} using the nuclear shell model, with an effective axial coupling constant $g_A$=0.9 and a meson exchange current enhancement $\epsilon_{MEC}$=1.4. The two theoretical total spectra, shown in the inset of Figure~\ref{94Y} normalized to 100 decays, have been then convoluted with the response function of the $\beta$ spectrometer. The resulting $\beta$ spectra are compared in Figure~\ref{94Y}, concluding that even with the high energy threshold conditions reported here at CENBG the new $\beta$ spectrometer potentially allows us to distinguish a $\beta$ shape effect.

\section{Conclusions}\label{sec-5}

In this work we have presented a new electron detector for $\beta$-shape studies mainly focussed on reactor applications and based on a thick plastic detector in coincidence with a thin silicon detector. The first measurements with this detector have been performed with the high-energy resolution electron-beam spectrometer at CENBG. The mono-energetic electron spectra measured allowed us to validate the MC response function of the detector. We have carefully implemented the geometry of the detector and the mechanics of the set-up in the MC simulations with the help of CADMesh libraries. Overall good qualitative agreement has been found when comparing experimental and simulated spectra for mono-energetic electron beams in the range 0.6-1.8~MeV and for a $^{207}$Bi conversion electron source. No influence of the electron-scattering model employed in the simulations was found in the reproduction of our experimental energy-deposited spectra. The good control of the simulations reported here is needed in order to obtain the shape of the $\beta$ spectra from the deconvolution of measured spectra. We have proved the potential sensitivity of this new $\beta$ detector for distinguishing different theoretical shape factors $C(Z,W)$ for the case of the $\beta$ decay of a relevant fission fragment, which will be quantitatively investigated in follow-up measurements with radioactive beams. As regards future experiments, a vacuum chamber to contain two identical telescopes has been designed and constructed for experimental campaigns at the IGISOL laboratory, where the full assembly with two telescopes has been recently commissioned~\cite{I246} and we have carried out the first measurements of $\beta$ spectra of fission products important for reactor antineutrino calculations~\cite{Alcala_ND}.

\acknowledgments

This work has been supported by the CNRS challenge NEEDS and the associated NACRE project, the CNRS/IN2P3 PICS TAGS between Subatech and IFIC, and the CNRS/IN2P3 Master projects Jyv\"askyl\"a and OPALE. This work has also been supported by the Spanish Ministerio de Econom\'ia y Competitividad under Grants No. FPA2017-83946-C2-1-P and No. RTI2018-098868-B-I00, by the Spanish Ministerio de Ciencia e Innovaci\'on under Grant No. PID2029-104714GB-C21, and by the Generalitat Valenciana under Prometeo Grant CIPROM/2022/9. Authors acknowledge the financial support from the Ministerio de Ciencia e Innovaci\'on  with funding from the European Union NextGenerationEU and Generalitat Valenciana in the call Programa de Planes Complementarios de I+D+i (PRTR 2022) under Project DETCOM, reference ASFAE/2022/027. W.G. acknowledges the support of the U.K. Science and Technology Facilities Council grant ST/P005314. V.G. acknowledges the support of the National Science Center, Poland, under Contract No. 2019/35/D/ST2/02081. Authors would like to thank Christine Marquet and her group at CENBG for making the electron spectrometer available to us and for the support during the measurements. Authors would also like to thank Neil Clark from MICRON, Massimo Volpi from CAEN, David Etasse from FASTER and Paul Schotanus from Scionix for fruitful discussions, as well as Leendert Hayen for sharing the $\beta$-shape theoretical corrections with our collaboration. The technical support of the electronic and mechanical services of Subatech is also acknowledged.

\bibliographystyle{unsrtnat}
\bibliography{eShape_bib}

\end{document}